\documentclass[a4paper]{jpconf}
\usepackage{graphicx}
\begin{document}
\title{$t\bar{t}$ pair hadroproduction in association with a heavy boson at the NLO QCD accuracy + Parton Shower}

\author{M V Garzelli$^1$, A Kardos$^{2,3}$, C G Papadopoulos$^{4,5}$, 
Z Tr\'ocs\'anyi$^2$}

\address{$^1$ Laboratory for Astroparticle Physics, University of Nova Gorica, SI~5000 Nova Gorica, Slovenia}
\address{$^2$ Institute of Physics and MTA-DE Particle Physics Research
Group, University of Debrecen,
H~4010 Debrecen  P.O.Box 105, Hungary}
\address{$^3$ INFN, Sezione di Milano-Bicocca, I~20126 Milano, Italy}
\address{$^4$ Institute of Nuclear Physics, NCSR $\Delta\eta\mu${\it \'o}$\kappa\rho\iota\tau o \varsigma$, GR~15310 Athens, Greece}
\address{$^5$ CERN, TH-Unit, CH~1211 Geneva 23, Switzerland}

\ead{garzelli@mi.infn.it}

\begin{abstract}
The {\texttt{PowHel}} framework allows to make predictions of total
and differential cross-sections of multiparticle hadroproduction
processes at both NLO QCD accuracy and NLO QCD matched to Parton Shower,
on the basis of the interface between the
{\texttt{POWHEG-BOX}} and {\texttt{HELAC-NLO}} codes.
It has already been applied to study several processes involving a
$t\bar{t}$ pair in association with a third particle or hadronic jet.
Our most recent predictions concern $t\bar{t}V$ hadroproduction (with
$V$ = $W$ or $Z$), at both parton and hadron level, by considering
different decay channels (hadronic and leptonic) of the heavy
particles. In particular, we show the results of our phenomenological
analyses under the same system of cuts also recently adopted by the CMS
collaboration at LHC.
\end{abstract}

\vspace*{-2em}
\section{Introduction}

In recent years significant effort was devoted to increase the accuracy
of theoretical predictions for $p$-$p$ collision processes, in order to
match the experimental capabilities of the LHC detectors. One line of
development aims at providing refined automatic computations of
fixed-order total and differential cross-sections with increased accuracy,
leading to predictions of the final products of hard-scattering
processes at the parton level. A second line of development, strongly
tied to the first one, relies on the combination of these results with
shower algorithms, able to track the evolution of these final-state
particles to hadronization.  In practice, to realize the combination
between hard-scattering and parton shower, different matching
schemes have been proposed, depending of the order of the perturbative
expansion. At NLO accuracy two methods have dominated the literature in the last
ten years, the MC@NLO~\cite{Frixione:2002ik} and POWHEG~\cite{Nason:2004rx,Frixione:2007vw}. Here we make use of the latter, as 
implemented in the computer program {\texttt{POWHEG-BOX}}~\cite{Alioli:2010xd}.

The {\texttt{POWHEG-BOX}} provides an almost automatic and
process-independent matching framework yet it requires some
process-dependent input, that has to be specified by the user, process
by process. Thus, it offers a framework to implement the matching of
new processes, once one is able to provide all required input. In case
of processes of some complexity already at the parton level,
e.g.~multiparticle production, this input can be conveniently computed
by external codes. 

Our {\texttt{PowHel}} package uses the {\texttt{HELAC-NLO}}
\cite{Bevilacqua:2011xh} set of codes, to provide all matrix-elements
required as input by {\texttt{POWHEG-BOX}} for the computation of
cross-sections at the NLO + PS accuracy for $t\bar{t}$ pair
hadroproduction in association with a third particle. The output of
{\texttt{PowHel}} are Les Houches event (LHE) files \cite{Alwall:2006yp},
including events at the first radiation emission level.  Additional
radiation emissions can be simulated by further showering
these events using Shower Monte Carlo (SMC) programs, such as
{\texttt{PYTHIA}}~\cite{Sjostrand:2006za} and
{\texttt{HERWIG}}~\cite{Corcella:2002jc}. This can be done a-posteriori,
since all essential process and event information needed at this purpose
is already stored in the LHE files. Thus the full method can be exploited
to produce predictions at the NLO level, at the first radiation emission
level (LHE level), after parton shower (PS) and after PS~+~hadronization~+~hadron decay.

The total cross-sections as computed from LHE events are expected to have NLO accuracy, whereas the corresponding differential distributions have formally NLO accuracy up to higher order corrections, as LHE events also include the effect of Sudakov form factors embedded in the POWHEG formula. Therefore LHE predictions, as compared to NLO ones, are very useful to check the correctness of the implementation and to understand, as well, in which phase-space regions one could expect that the effects of higher-order corrections will sensibly modify the NLO distributions. On the other hand, predictions after the full SMC chain are the closest to the experimental results, allowing for an immediate comparison with the data collected in the detectors. Actually, in many experiments reconstruction procedures have been developed in order to reconstruct the intermediate unstable particles, like top quarks and heavy bosons, from which the final-state products are originated (bottom-up approach). We would like to emphasize however, that, once full high-accuracy simulations from the parton level down to the hadron level (top-down approach) are available, like in the case of {\texttt{PowHel}} + SMC, this reconstruction is unnecessary.

In this paper we concentrate on the case of $t\bar{t}$-pair
hadroproduction in association with an heavy gauge boson, $t\bar{t} W$
and $t\bar{t}Z$, that we have recently studied extensively in
Ref.~\cite{Garzelli:2012bn,Garzelli:2011is}, leaving the reader
interested in our predictions for other $t\bar{t}$-X associated processes,
like $t\bar{t}j$, $t\bar{t}H$ and $t\bar{t}A$, to 
previous works including {\texttt{PowHel}} predictions~\cite{Kardos:2011qa,Garzelli:2011vp,Dittmaier:2012vm}.

\section{Results}

$t\bar{t}Z$ and $t\bar{t}W^{\pm}$ LO and NLO inclusive cross-sections as
predicted by {\texttt{PowHel}} were already presented in
Ref.~\cite{Garzelli:2012bn,Garzelli:2011is}. Here we emphasize that
these predictions follow extensive checks and comparisons to the
results of previous works presented by different authors in
Ref.~\cite{Lazopoulos:2008de,Campbell:2012dh}, on the basis of
different methods. Further predictions concerning several differential
distributions at NLO and at NLO + PS accuracy were included, as for
$t\bar{t}Z$ hadroproduction, for the first time in
Ref.~\cite{Kardos:2011na} and~\cite{Garzelli:2011is}, respectively,
whereas, as for $t\bar{t}W$, in Ref.~\cite{Garzelli:2012bn}. 

Our predictions use the following parameters: $m_t=$ 172.5 GeV,
$m_W$ = 80.385 GeV, $m_Z$~=~91.1876 GeV, $\sin^2 \theta_C$ = 0.049284,
$\mu_R =\mu_F =m_t + m_V/2$, where $m_V$ is the mass of the vector boson,
and the {\texttt{CTEQ6.6M}} PDF set available with the LHAPDF interface, 
with a 2-loop running $\alpha_S$.
We adopted the SMC versions {\texttt{PYTHIA-6.4.26}} and
{\texttt{HERWIG-6.520}}. We enforced stability of $\pi^0$ in both,
whereas $\mu$ stability was enforced in {\texttt{HERWIG}}, to match the
default in {\texttt{PYTHIA}}. All other hadrons and leptons were assumed to
be stable or to decay with B.R. according to the default implementation
of each SMC.

When making predictions at NLO + PS accuracy, we concentrated on two
different channels, the trilepton and the same-sign dilepton ones, as
also recently exploited by the CMS collaboration in
Ref.~\cite{CMS:2012pwa}. Essentially, the trilepton channel selection
aims at isolating the $Z$, decaying in $\ell^+ \ell^-$ pairs, together
with the $t\bar{t}$ pair decaying semileptonically. The selection of
the dilepton channel aims at isolating the $W^{\pm}$, decaying in
$\ell^{\pm} {^(}\bar{\nu}^)_\ell$, again with the $t\bar{t}$ pair
decaying semileptonically. The list of cuts we adopted for this study
is extensively detailed in Ref.~\cite{Garzelli:2012bn}.

\begin{figure}[t!]
\includegraphics[width=7.5cm]{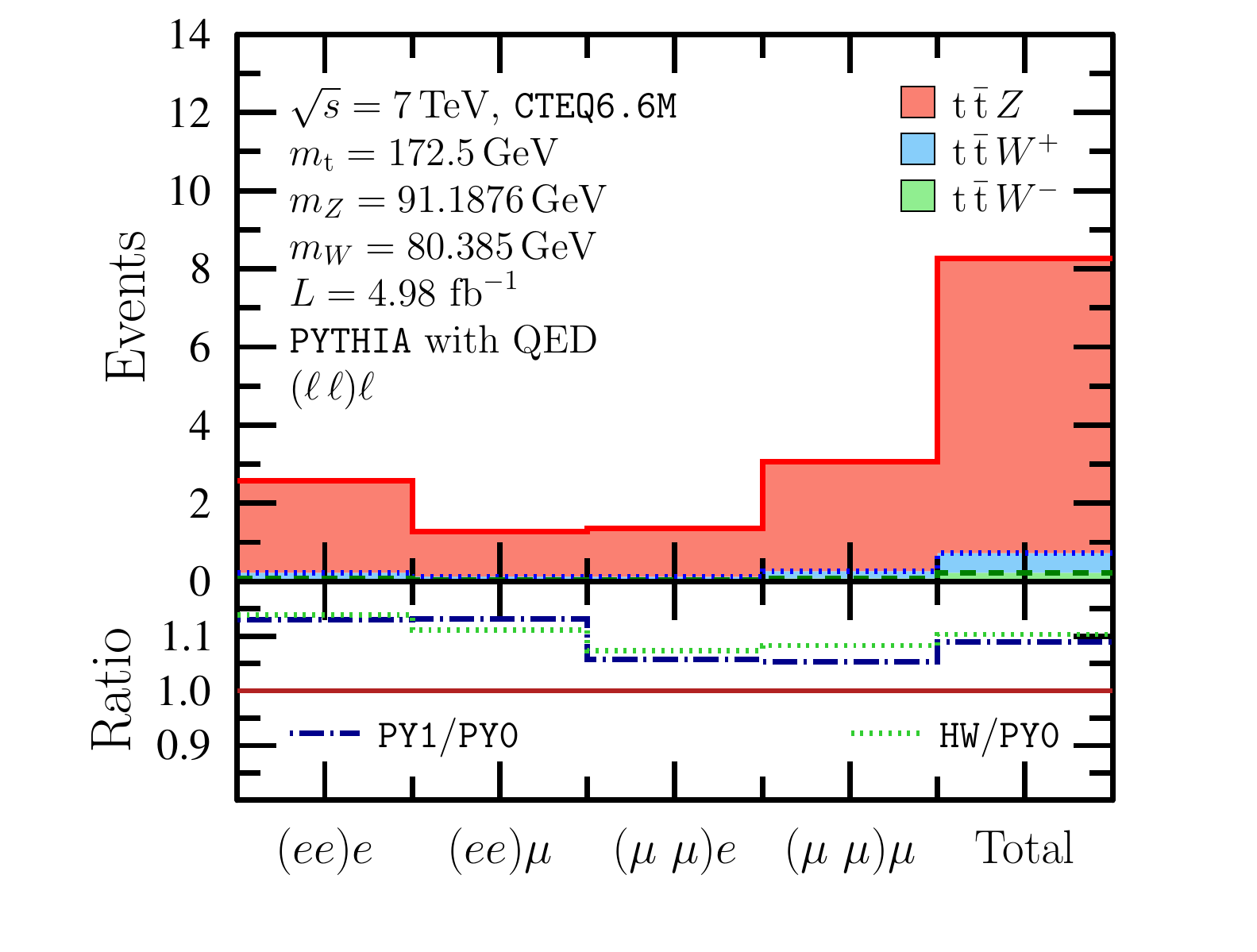}
\includegraphics[width=7.5cm]{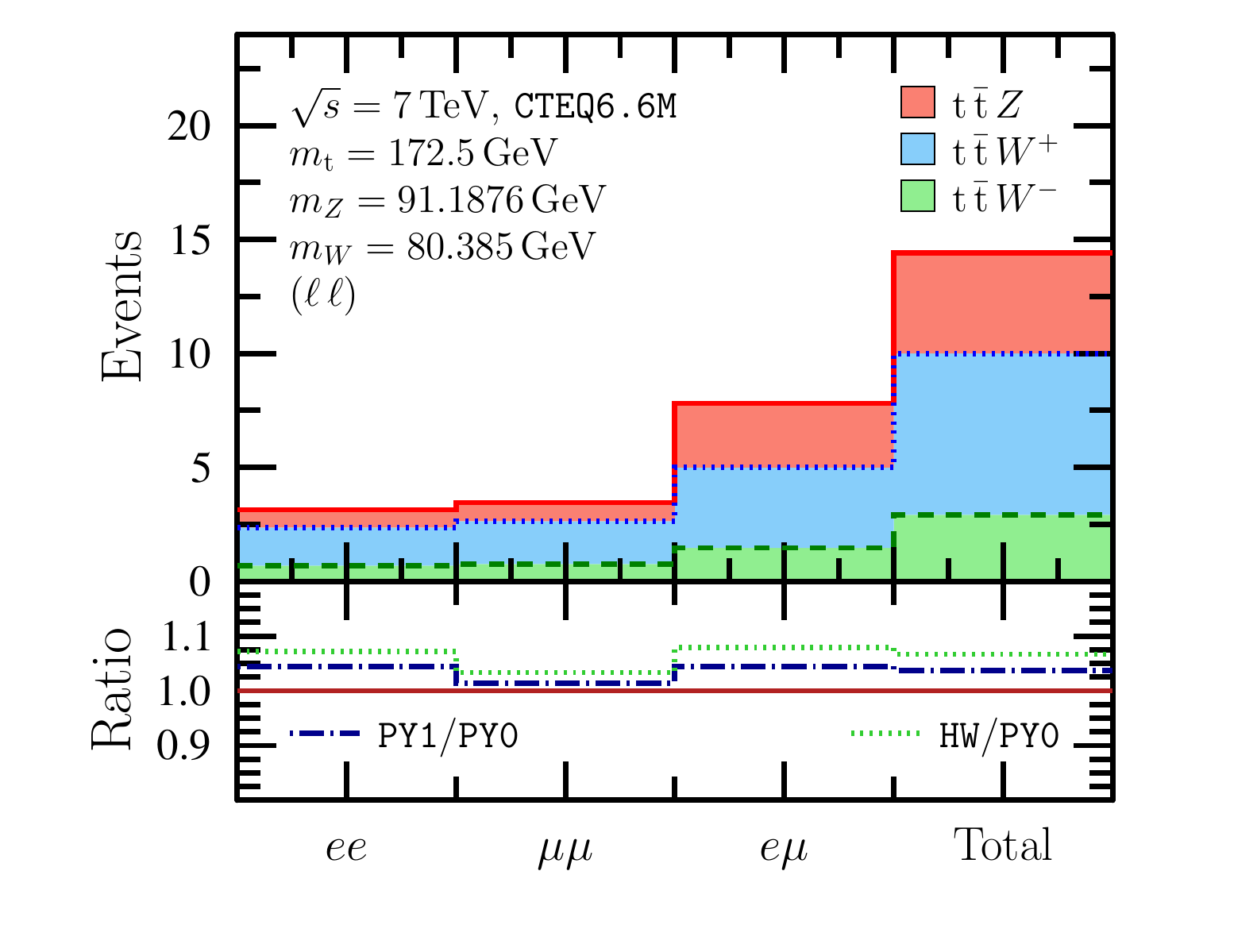}
\caption{\label{fig1}
Number of events in a) the trilepton and b) the dilepton channels at
the $\sqrt{s} = 7$ TeV LHC, as predicted by \texttt{PowHel}
+~\texttt{PYTHIA}, for an integrated luminosity $L = 4.98$\,fb $^{-1}$.
The contribution in all possible channels are shown separately, as well
as their sum in the last bin of each panel.  The contributions due to
$t\bar{t}Z$, $t\bar{t}W^+$ and $t\bar{t}W^-$ are cumulated one over the
other. To be compared with the experimental data in Fig.~4 and~6 of the
CMS technical report~\cite{CMS:2012pwa}, respectively.
In the lower inset the ratios between cumulative
results using different SMC (\texttt{HW}/\texttt{PY}) and between 
cumulative results obtained by neglecting and including photon radiation
from leptons (\texttt{PY1}/\texttt{PY0}) are also shown.}
\end{figure}

\begin{figure}[t!]
\includegraphics[width=7.5cm]{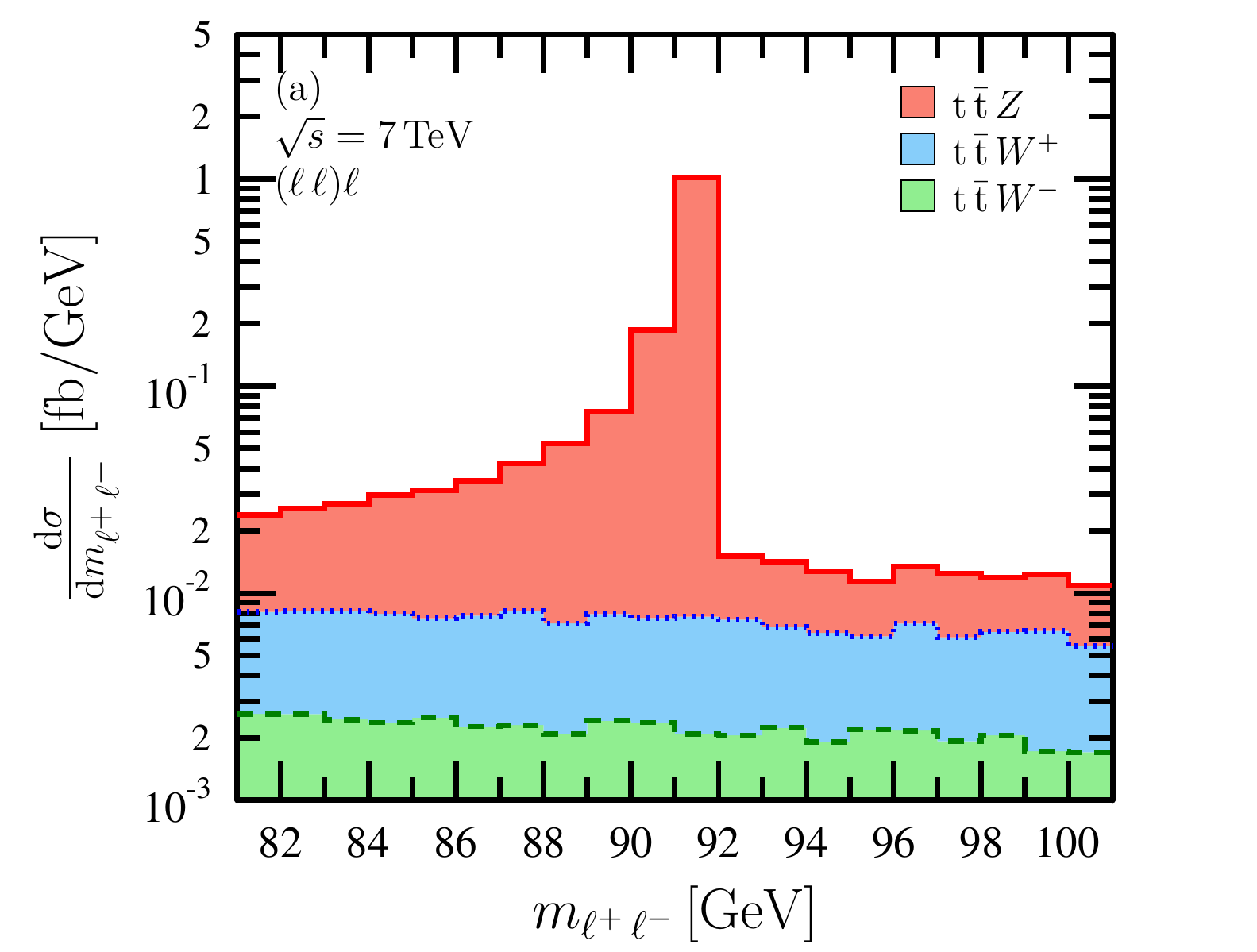}
\includegraphics[width=7.5cm]{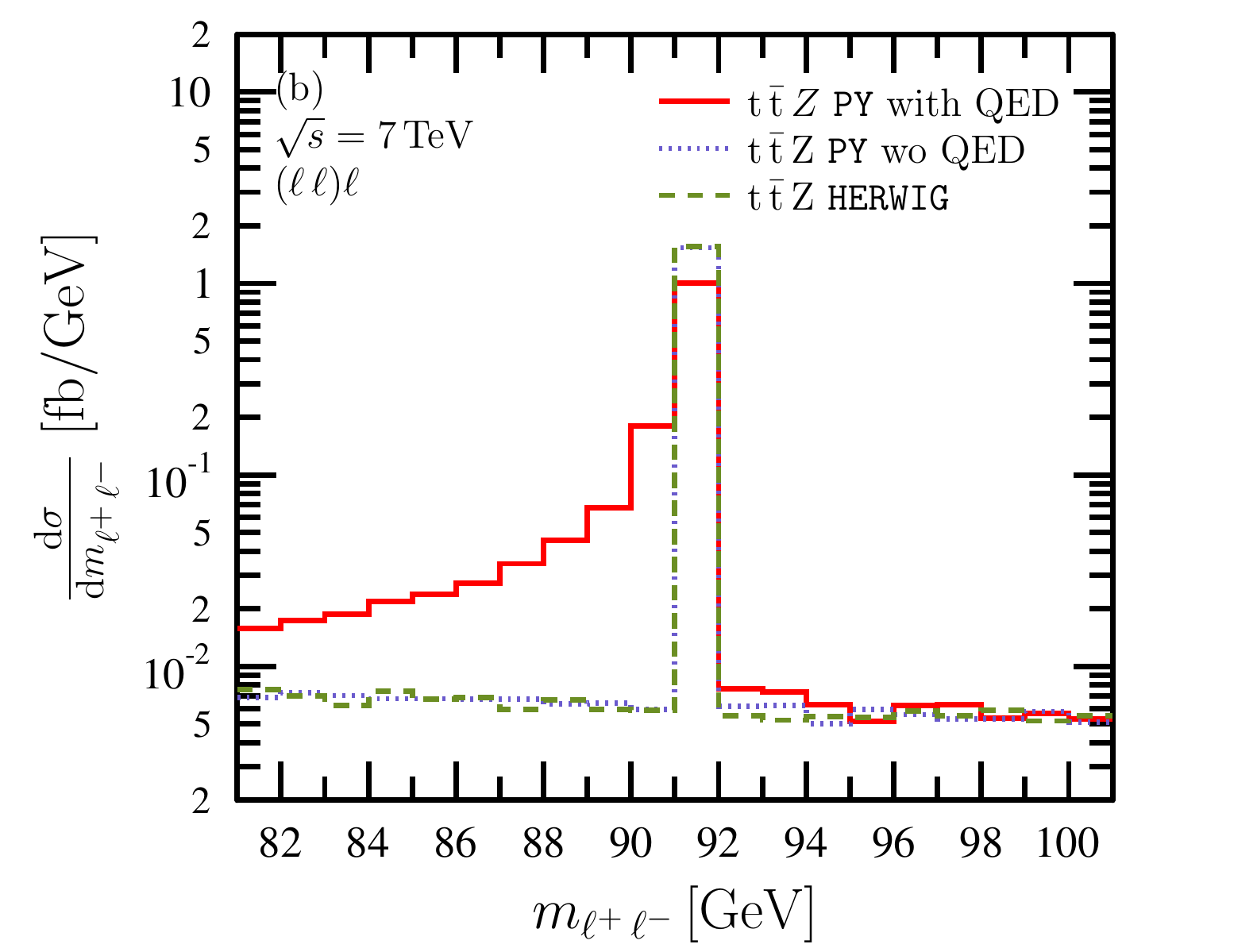}
\caption{\label{fig2} Invariant mass of the Z reconstructed from
same-flavour ($\ell^+$, $\ell^-$) pairs after the trilepton analysis,
as obtained by \texttt{PowHel} +~\texttt{PYTHIA} at the $\sqrt{s} =$ 7
TeV LHC: in panel a) the results corresponding to the different
processes  $t\bar{t}Z$, $t\bar{t}W^+$ and $t\bar{t}W^-$ are cumulated one
over the other, whereas in panel b) distributions obtained by using
different SMC (\texttt{PYTHIA}, \texttt{HERWIG} and \texttt{PYTHIA}
without photon radiation from leptons) are also shown, limited to
$t\bar{t}Z$.}
\vspace*{-1em}
\end{figure}

In the trilepton channel, requesting the invariant mass of a pair of
opposite-charge isolated leptons to be close to the $Z$ mass, the
contribution of $t\bar{t}W^+$ and $t\bar{t}W^-$ is suppressed with
respect to the contribution of $t\bar{t}Z$, as shown in
Fig.~\ref{fig1}.a.  The invariant mass of the reconstructed $Z$ after
cuts is shown in Fig.~\ref{fig2}.a, by cumulating the results of the
three processes. Pairs of opposite charge leptons, emitted by heavy
particle decays, turned out to be quite insensitive to SMC effects,
allowing for a good reconstruction of heavy objects. This opens good
perspectives for the study, in the trilepton channel, of even other
neutral boson candidates, like an exotic $Z^\prime$. Even if further
photon emissions from the emitted leptons smear the $Z$ peak in the
($\ell^+$, $\ell^-$) invariant mass distribution, this does not
disappear, as shown in Fig.~\ref{fig2}.b. 
In the experimental event
reconstruction electrons are often corrected for photon emission
effects, but this is not always the case of muons. Thus, one could
treat the two cases separately even in the shower simulation, since it
is not always true that photon emission from muons leads to negligible
effects. 


The reconstruction of charged heavy objects like a $W$ boson or a
$t$-quark is complicated because the measurement of missing energy,
attributed to heavy particle decay, has large uncertainties.  As a
consequence, we did not introduce any explicit cut to distinguish
$t\bar{t}W$ from $t\bar{t}Z$ contributions in the analysis in the
same-sign dilepton channel. Thus the contribution of the $t\bar{t}Z$
process is not negligible with respect to that of the $t\bar{t}W^\pm$
ones, so all three processes contribute in a consistent way to the total
amount of observed events, as shown in Fig.~\ref{fig1}.b.

The total cumulative cross-sections after full SMC, including the
contributions of $t\bar{t}Z$, $t\bar{t}W^+$ and $t\bar{t}W^-$, for all
different lepton combinations in the trilepton and dilepton decay
channels, at 7 and 8 TeV, are reported in Tables~\ref{table1}.a and~\ref{table1}.b. In the last column of each table, the ratio of the
results at 7 and 8 TeV for each single channel is also reported showing
an almost constant value. The slightly different average value of this
ratio from all trilepton channels, as compared to the average from all
dilepton channels, is related to the different set of cuts.

\begin{table}
\centering
\begin {tabular}{@{}cccc@{}}
\hline\hline
channel & {\texttt{PowHel+SMC}} & {\texttt{PowHel+SMC}} & ratio\\
$\,$ & 7 TeV (fb)  & 8 TeV (fb) & $\,$ \\
\hline
($e$,$e$)$e$ & 0.516 & 0.782 & 1.515\\
($e$,$e$)$\mu$ & 0.255 & 0.388 & 1.521\\
($\mu$,$\mu$)$e$ & 0.273 & 0.420 & 1.538\\
($\mu$,$\mu$)$\mu$ & 0.613 & 0.934 & 1.523\\
\hline
total & 1.658 & 2.524 & 1.522\\
\hline
\hline
\end{tabular}
\hfill
\begin {tabular}{@{}cccc@{}}
\hline\hline
channel & {\texttt{PowHel+SMC}} & {\texttt{PowHel+SMC}} & ratio\\
$\,$ & 7 TeV (fb)  & 8 TeV (fb) & $\,$ \\
\hline
($e$,$e$) & 0.631 & 0.907 & 1.437\\
($e$,$\mu$) & 0.694 & 0.991 & 1.428\\
($\mu$,$\mu$) & 1.569 & 2.289 & 1.459\\
\hline
total & 2.894 & 4.187 & 1.446\\
\hline
\hline
\end{tabular}
\caption{Cumulative cross-sections at $\sqrt{s} =$ 7 and 8 TeV, as predicted by {\texttt{PowHel+PYTHIA}} in a) all trilepton and b) all same-sign dilepton channels. The statistical error is always $<$ 1\%.}
\label{table1}
\end{table}

The $\sqrt{s} =$ 7 and 8 TeV {\texttt{PowHel}} events used for these
analyses are freely available on the web at {\texttt
{http://www.grid.kfki.hu/twiki/bin/view/DbTheory/WebHome}}.

\section*{Acknowledgments}
{\small We are grateful to the TOP-2012 Workshop organizers and partecipants for many interesting discussions. This work was supported by the LHCPhenoNet network PITN-GA-2010-264564, the Hungarian Scientific Research Fund grant K-101482 and the Slovenian Ministery of Work under the AD-FUTURA program.}

\section*{References}
\providecommand{\newblock}{}

\end{document}